\documentclass[twocolumn,aps,pre,superscriptaddress]{revtex4-2}
\usepackage{float}
\usepackage{amsmath}
\usepackage{amssymb}
\usepackage{graphicx}
\usepackage[dvipsnames]{xcolor}
\usepackage{comment}
\usepackage{subfig}
\usepackage[utf8]{inputenc}
\usepackage{hyperref}
\usepackage{epstopdf}
\usepackage{siunitx}
\usepackage{ragged2e}
\usepackage{xcolor}

\graphicspath{{./Figures/}}

\begin{document}

\title{Semiflexible Ring Polymers on Active Motor Beds: Nonequilibrium Dynamics and Conformations}

\author{Sandip Roy}
\affiliation{Indian Institute of Science Education and Research Mohali,
Knowledge City, Sector 81, SAS Nagar 140306, Punjab, India}
\affiliation{Tata Institute of Fundamental Research Hyderabad,
36/P, Gopanpally Village, Serilingampally Mandal,
Ranga Reddy District, Hyderabad, Telangana 500046, India}

\author{Abhishek Chaudhuri}
\affiliation{Indian Institute of Science Education and Research Mohali,
Knowledge City, Sector 81, SAS Nagar 140306, Punjab, India}

\author{Anil Kumar Dasanna}
\email[]{adasanna@iisermohali.ac.in}
\affiliation{Indian Institute of Science Education and Research Mohali,
Knowledge City, Sector 81, SAS Nagar 140306, Punjab, India}
\date{\today}

\begin{abstract}
A semiflexible ring polymer on a motor-protein bed exhibits activity- and processivity-dependent rotational and conformational dynamics that are not captured by linear-chain behavior. Using coarse-grained Langevin simulations with bending elasticity, excluded-volume interactions, and stochastic motor attachment, stepping, and detachment, we vary activity (P\'eclet number), motor processivity, and chain stiffness to map the nonequilibrium response. The mean-squared displacement shows crossover dynamics, with semiflexible rings displaying subdiffusive-to-diffusive behavior at low activity and an intermediate ballistic regime at higher activity, while increasing flexibility shifts the short-time response toward a Rouse-like limit. Diameter autocorrelations exhibit damped oscillations associated with coherent rotation; the rotational frequency increases with activity and processivity, whereas the decorrelation time is non-monotonic at high processivity. Fourier mode analysis identifies competition between the radius ($k=0$) and elliptic ($k=2$) modes as the origin of the non-monotonic asphericity.
\end{abstract}

\maketitle

\section{Introduction}

The cytoskeleton is a dynamic and interconnected filament network that governs the structural organization, mechanics, and transport functions of eukaryotic cells. It comprises three major filament systems: actin-based microfilaments ( diameter $\sim 7$ nm), intermediate filaments ( diameter $\sim 10$ nm), and microtubules ( diameter $\sim 11$--$15$ nm). Continuous polymerization and depolymerization of these filaments enable cells to reorganize their architecture and mechanical response in a highly adaptive manner~\cite{alberts1989molecular}.

Motor proteins (MPs) are active molecular machines that interact with cytoskeletal filaments and convert chemical energy from ATP hydrolysis into directed mechanical motion. Distinct motor families, such as myosins and kinesins, differ in filament selectivity, stepping kinetics, and biological function, and are central to intracellular transport and force generation. Their roles span organelle transport (including mitochondria~\cite{kruppa2021motor}, Golgi stacks~\cite{burkhardt1998role}, and vesicles~\cite{hirokawa2009kinesin}), as well as larger-scale cytoskeletal processes such as muscle contraction~\cite{veigel1999motor}, ciliary and flagellar beating~\cite{lindemann2010flagellar,riedel2007molecular}, and cell division~\cite{scholey2003cell}.

Because motor proteins continuously consume energy, motor--filament assemblies are intrinsically nonequilibrium systems that break detailed balance~\cite{julicher1997modeling} and do not satisfy equilibrium fluctuation--dissipation relations. In vitro motility assays, in which actin filaments or microtubules move on surfaces coated with immobilized motors, provide a controlled platform to probe the emergent physics of such active systems~\cite{rogers2004motility,scholey1993motility,vale1985identification,uyeda1990myosin,harada1987sliding,bourdieu1995spiral,schaller2010polar}. Experiments and simulations in these settings have revealed a broad range of collective and nonequilibrium filament dynamics, including gliding, spiral formation, oscillations, and large-scale swirling states~\cite{duke1995gliding,liverpool2001viscoelasticity,ghosh2014dynamics,harder2014activity,kaiser2014unusual,laskar2015brownian,eisenstecken2016conformational,isele2015self,isele2016dynamics,winkler2017active,lowen2018active,sarkar2014ring,chelakkot2014flagellar,loi2011non,jayaraman2012autonomous,jiang2014motion,eisenstecken2017internal,gupta2019morphological,yadav2024wave,roy2025confinement,roy2025spirals,khosravanizadeh2025dynamic}.

Most work in motility-assay settings has focused on linear filaments. By contrast, closed or ring polymers remain comparatively less explored, despite their relevance to circular DNA, cortical actin-ring-like structures, and coarse-grained models of vesicle- or membrane-like objects. Ring topology qualitatively changes the dynamics by eliminating free ends and coupling local forcing to global rotational and conformational modes. Recent studies on active ring polymers have begun to uncover topology-specific nonequilibrium behaviors, including contour-following motion, activity-induced conformational changes, and excluded-volume effects~\cite{winkler2020physics,zhu2024non,theeyancheri2023active,kumar2023local,teixeira2021single,goychuk2024delayed,mousavi2019active,locatelli2021activity,philipps2022dynamics,lamura2024excluded}. These dynamics also share features with passive and active ring-like objects in flow, such as tank-treading and contour-aligned motion in sheared rings and vesicles~\cite{liebetreu2018trefoil,chen2013tumbling,winkler2024active,keller1982motion,noguchi2004fluid}.Motility assays serve as test-beds for a number of such studies as they naturally provide an active setting which can be tuned. They provide well-controlled biomimetic platforms in which this simplified limit can be explored.


We perform Langevin dynamics simulations of a semiflexible ring polymer with bending elasticity and excluded-volume interactions in a motility assay.
The system is controlled by three key dimensionless parameters: the P\'eclet number ($Pe$), which sets the motor activity; the motor processivity ($\Omega$), which quantifies the balance of attachment and detachment kinetics; and the persistence ratio ($l_p/L$), which determines polymer stiffness. We show that ring topology, motor kinetics, and stiffness together produce coupled transport, rotational, and conformational dynamics. The mean-squared displacement exhibits activity-dependent crossover behavior, including subdiffusive, ballistic, and diffusive regimes depending on stiffness and activity. Diameter autocorrelations reveal damped oscillations associated with coherent rotation, with a rotational frequency that increases with activity and processivity but is only weakly affected by stiffness, while the corresponding decorrelation time becomes non-monotonic at high processivity. Conformational analysis further shows pronounced non-monotonicity in ring asphericity for stiff rings at high processivity. A Fourier mode decomposition provides a mechanistic interpretation of this behavior in terms of competition between the radius ($k=0$) and elliptic ($k=2$) modes. Finally, direct measurements of angular velocity provide an independent kinematic diagnostic of the rotational dynamics inferred from diameter autocorrelations.
We note that while our findings are most directly relevant to simplified in vitro motility-assay settings, they also provide a minimal reference point for thinking about how motor activity, filament elasticity, and closed topology can interact in more complex biological ring-like assemblies.

The paper is organized as follows. Section~\ref{sec:Model} introduces the polymer model and motor-protein dynamics, along with simulation details. Section~\ref{sec:Results} presents the results: Sec.~\ref{sec:Dynamics} focuses on dynamical properties, including mean-squared displacement, diameter autocorrelations, and angular velocity, while Sec.~\ref{sec:Shape} examines conformational properties, including tangent--tangent correlations, ring size, asphericity, and mode analysis. We conclude in ~\ref{sec:SummaryOutlook} with a summary and outlook.

\begin{figure}[h!tbp]
\includegraphics[width=\columnwidth]{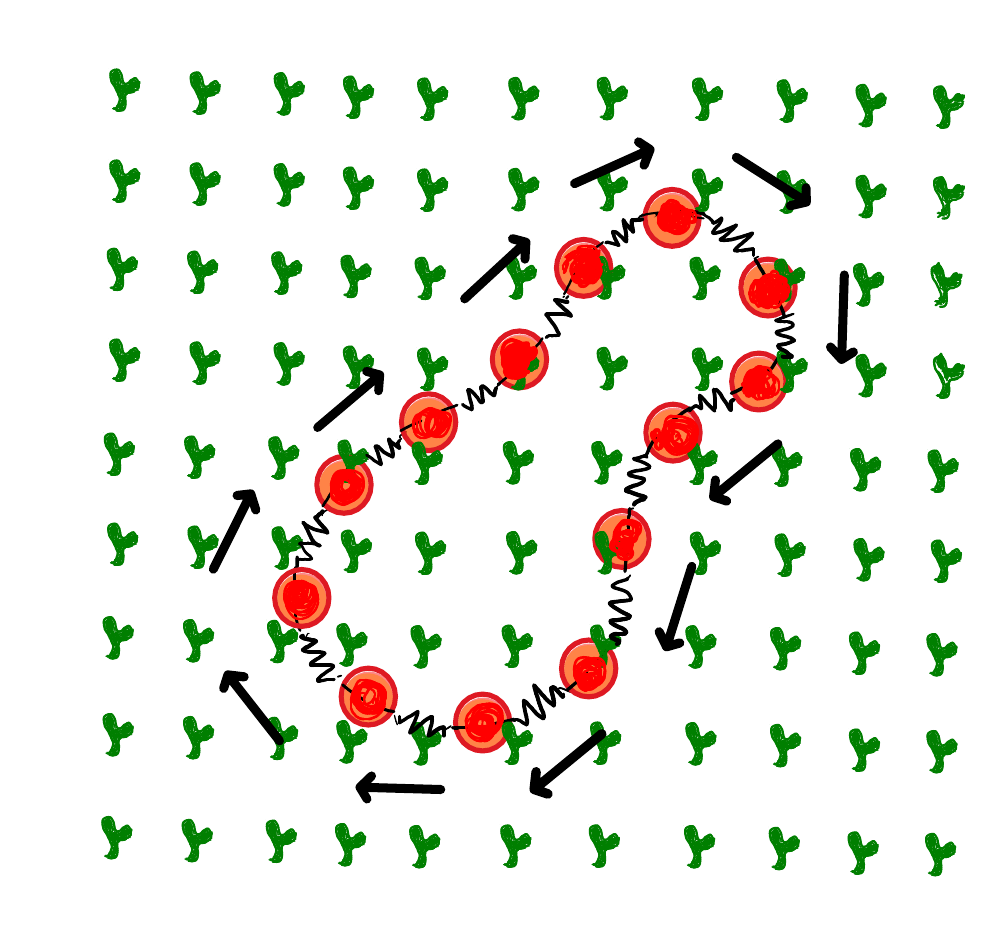}
\caption{\justifying
Schematic 2D representation of the model system. The polymer is modeled as a bead--spring chain (red) placed on a lattice of motor proteins (green). Motor heads attach when a polymer segment enters their capture radius and subsequently step along the contour. Blue arrows indicate the direction of ring rotation; a clockwise ring rotation corresponds to motors stepping anticlockwise along the contour.
}
\label{fig:Schematic}
\end{figure}

\section{Model and Simulation}
\label{sec:Model}

We extend our earlier motility-assay models for semiflexible linear polymers~\cite{gupta2019morphological,shee2021semiflexible} to a closed (ring) polymer geometry. In this section, we describe the polymer Hamiltonian, the motor-protein kinetics, and the numerical integration scheme used in the simulations.

\subsection{Polymer Model}
\label{sec:Model:Poly}

The polymer is modeled as a closed bead--spring chain of $N$ monomers with bending rigidity and excluded-volume interactions in two dimensions. Monomer positions are denoted by $\{\vec r_i\}_{i=1}^N$, and ring topology is enforced by periodic indexing, i.e., $\vec r_{N+1}\equiv \vec r_1$ and, more generally, indices are understood modulo $N$. The contour length (perimeter) of the ring is $L = N r_0$, where $r_0$ is the equilibrium bond length (Fig.~\ref{fig:Schematic}).

We define bond vectors and local tangents as $\vec b_i = \vec r_{i+1}-\vec r_i, \hat t_i = {\vec b_i}/{|\vec b_i|}$, with $i=1,\dots,N$ (periodic indexing implied). The total polymer energy is ${\cal E} = {\cal E}_s + {\cal E}_b + {\cal E}_{\rm WCA}$. The stretching energy penalizes bond-length fluctuations:
\begin{equation}
{\cal E}_s = \sum_{i=1}^{N} \frac{K_{\rm s}}{2}\,\big(|\vec b_i|-r_0\big)^2.
\end{equation}
In practice, $K_{\rm s}$ is chosen sufficiently large so that the bonds remain nearly inextensible. The bending energy penalizes local curvature:
\begin{equation}
{\cal E}_b = \sum_{i=1}^{N} \frac{\kappa}{2r_0}\,\big\lVert \hat t_{i+1}-\hat t_i\big\rVert^2,
\end{equation}
where $\kappa$ is the bending modulus. The corresponding persistence length is $l_p = 2\kappa/k_BT$ in two dimensions.

Non-bonded monomers interact through a purely repulsive Weeks--Chandler--Andersen (WCA) potential, ${\cal E}_{\rm WCA} = \sum_{i<j}' U_{\rm WCA}(r_{ij})$, where $r_{ij}=|\vec r_i-\vec r_j|$ and the prime indicates that directly bonded nearest neighbors are excluded from the WCA sum. The pair potential is
\begin{equation}
U_{\rm WCA}(r)=
\nonumber
\begin{cases}
4\epsilon \left[\left(\dfrac{\sigma}{r}\right)^{12}-\left(\dfrac{\sigma}{r}\right)^6+\dfrac14\right], & r<2^{1/6}\sigma,\\[6pt]
0, & r\ge 2^{1/6}\sigma.
\end{cases}
\end{equation}

We perform Langevin dynamics simulations in a periodic square domain of size $L_b = 100\,\sigma$.
\begin{equation}
m\ddot{\vec r}_i = -\zeta \dot{\vec r}_i - \nabla_i {\cal E} + \vec F^{\,\rm motor}_i + \boldsymbol{\xi}_i(t),
\label{eq:langevin_bead}
\end{equation}
where $\zeta$ is the monomer friction coefficient, $\vec F^{\,\rm motor}_i$ is the net force exerted by bound motors on bead $i$, and $\boldsymbol{\xi}_i(t)$ is Gaussian thermal noise with
\begin{equation}
\nonumber
\langle \xi_{i\alpha}(t)\rangle = 0,\qquad
\langle \xi_{i\alpha}(t)\xi_{j\beta}(t')\rangle
=2\zeta k_BT\,\delta_{ij}\delta_{\alpha\beta}\delta(t-t').
\end{equation}
Here $\alpha,\beta\in\{x,y\}$ denote Cartesian components. 

\subsection{Motor Protein Dynamics}
\label{sec:Model:MP}

The motors are placed on a square lattice of area fraction $\rho$ on the substrate (Fig.~\ref{fig:Schematic}). Each motor consists of an anchored tail and an active head. A motor can bind stochastically to a polymer segment if the head lies within a capture radius $r_c$ of the segment, with attachment rate $\omega_{\rm on}$. Once bound, the motor head moves unidirectionally along the local filament tangent, mimicking myosin-like stepping. Because the polymer is a closed ring, there are no filament ends, and motors can continue stepping around the contour as long as they remain attached.

A bound motor behaves as a spring with force $\vec f_\ell = -k_m \Delta \vec r$, where $k_m$ is the motor stiffness and $\Delta \vec r$ is the displacement of the motor head from its anchored tail. This is the standard small-extension approximation for coarse grained motor assays. If the motor is attached to a point in the segment between the beads $i$ and $i+1$, the motor force is distributed to the two beads using a linear lever rule. Specifically, if the attachment point is a fractional distance $\lambda\in[0,1]$ from the bead $i$ along the segment, then $\vec F^{\,\rm motor}_i \to \vec F^{\,\rm motor}_i + (1-\lambda)\vec f_\ell, \vec F^{\,\rm motor}_{i+1} \to \vec F^{\,\rm motor}_{i+1} + \lambda \vec f_\ell$. Motor stepping is load-dependent. The active stepping velocity along the local tangent is taken as
\begin{equation}
v_t^a(f_t)=\frac{v_0}{1+d_0\exp(f_t/f_s)},
\end{equation}
where $v_0$ is the unloaded motor speed, $f_s$ is the stall force, $d_0$ is a force-sensitivity parameter, and $f_t = -\vec f_\ell\cdot \hat t$ is the tangential load opposing stepping. This relation was previously derived from a coarse-grained ATP-hydrolysis/Michaelis–Menten description and calibrated to kinesin force–velocity data with $d_0 = 0.012$ and $f_s \simeq 2$ pN~\cite{chaudhuri2016forced,schnitzer2000force}. Motor detachment is modeled by Bell's law~\cite{bell1978models},
\begin{equation}
\omega_{\rm off}(f_l) = \omega_0 \exp(f_\ell/f_d),
\end{equation}
where $f_\ell = |\vec f_\ell|$, $\omega_0$ is the zero-load detachment rate and $f_d$ is the characteristic detachment force. The corresponding effective (load-dependent) processivity is
\begin{equation}
\Omega(f_\ell)=\frac{\omega_{\rm on}}
{\omega_{\rm on}+\omega_0\exp(f_\ell/f_d)},
\end{equation}
which reduces as the motor load increases. In the present study, we set $f_d = f_s$ as a simplifying baseline choice.

In the numerical implementation, motor attachment and detachment are updated stochastically at each timestep using the rates $\omega_{\rm on}$ and $\omega_{\rm off}$, respectively (with probabilities $\omega_{\rm on}\Delta t$ and $\omega_{\rm off}\Delta t$ for sufficiently small $\Delta t$), while the bound motor head position is advanced continuously along the contour according to the load-dependent velocity $v_t^a$.

\subsection{Simulation Parameters and Units}
\label{sec:Simu}

Unless otherwise stated, we simulate rings of $N=50$ monomers with $r_0=\sigma=1$, so that $L=N r_0$. The energy scale is set by $k_BT=\epsilon=1$. The monomer friction coefficient is $\zeta=10$, and the monomer mass is $m=1$. The persistence ratio $l_p/L$ is used to tune polymer stiffness, with $l_p/L=0.3$ corresponding to a semiflexible ring. The motor parameters are chosen as $f_s=2k_BT/\sigma$, $f_d=f_s$, $k_m=K_s$ and the motor density is $\rho \simeq 3.8/\sigma^2$, which corresponds to an average inter-motor spacing of approximately $0.5\,\sigma$. The motor capture radius is $r_c=0.5~\sigma$. Motor activity is quantified by the P\'eclet number $Pe = {v_0 L^2}/{D\sigma}$, where $D=k_BT/\zeta$ is the monomer diffusion coefficient.  Time is nondimensionalized by the passive ring relaxation timescale: $\tau={L^3\zeta}/{24\pi^2 k_BT}$~\cite{philipps2022dynamics}. Simulations are run for $2\times 10^8$ timesteps with timestep $\delta t \approx 1.6\times 10^{-8}\tau$, and the first $10^8$ steps are discarded as equilibration. The initial polymer conformation is a circle of diameter $\simeq L/\pi$. All reported quantities are averaged over 20 independent realizations, with error bars representing the standard error of the mean. We define the bare processivity rate as $\Omega = \omega_{on}/(\omega_{on}+\omega_{0})$. All parameters used in the simulations are summarized in Table~\ref{table}.

\begin{table}[h!]
\small
  \caption{Simulation parameters.}
  \label{tbl:example1}
  \begin{tabular*}{0.48\textwidth}{@{\extracolsep{\fill}}lll}
    \hline
    Parameter & Definition & Value \\
    \hline
    $m$ & Mass of each bead & 1 \\
    $N$ & Number of beads & 50 \\
    $\sigma$ & Effective bead diameter & 1 \\
    $r_0$ & Bond length & 1 \\
    $k_BT$ & Energy scale & 1 \\
    $r_c$ & Motor capture radius & $0.5\,\sigma$ \\
    $K_\textrm{s}$ & Bond stiffness & $100\,k_BT/\sigma^2$ \\
    $\rho$ & Motor density & $3.8\,\sigma^{-2}$ \\
    $f_d$ & Detachment force & $2\,k_BT/\sigma$ \\
    $f_s$ & Stall force & $2\,k_BT/\sigma$ \\
    $d_0$ & Force sensitivity & 0.012 \\
    $\zeta$ & Bead friction coefficient & 10 \\
    $\zeta_{MP}$ & Motor head friction & $1\,\zeta$ \\
    $k_m$ & Motor stiffness & $100\,k_BT/\sigma^2$ \\  
    $\Omega$ & Processivity range & 0.1–0.9 \\
    \hline
  \end{tabular*}
  \label{table}
\end{table}

\begin{figure*}[!htb]
\includegraphics[width=\textwidth]{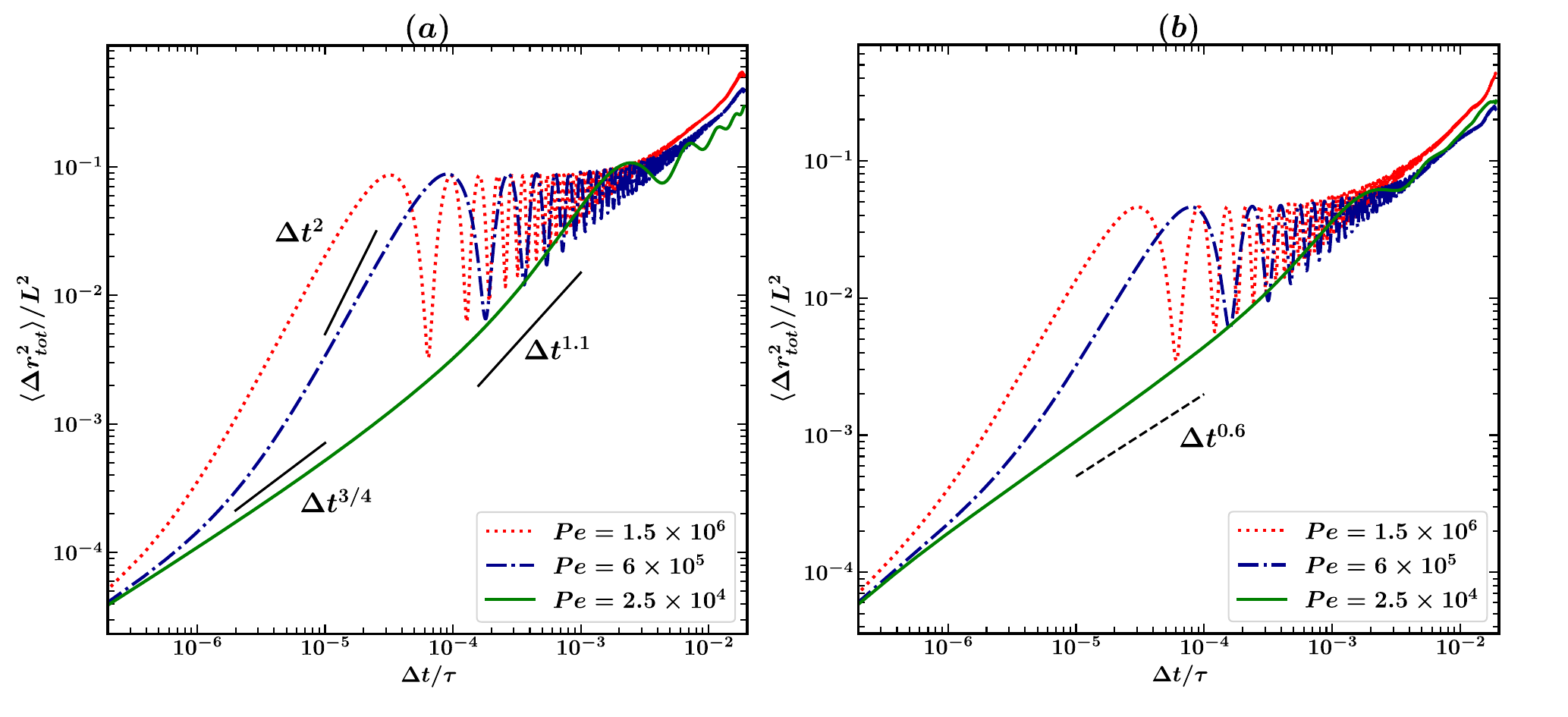}
\caption{\justifying
Mean-squared displacement (MSD) of a semiflexible polymer ring on a motor-protein bed at fixed bare processivity \(\Omega=0.5\). 
Panels show \(\langle \Delta r^2_{\mathrm{tot}}(\Delta t)\rangle/L^2\) for different \(Pe\) at (a) \(l_p/L=0.5\) and (b) \(l_p/L=0.03\). 
The MSD displays the expected active-polymer crossover regimes, while the oscillatory modulations indicate the onset of coherent rotational motion.
}
\label{polyRing_on_MP_MSD_VaryPersistence}
\end{figure*}

\section{Results}
\label{sec:Results}

\subsection{Dynamics of Polymer Ring}
\label{sec:Dynamics}

\subsubsection{Transport Behavior}

The mean-squared displacement (MSD) of the ring contains contributions from center-of-mass (COM) motion and internal conformational fluctuations. We define the total MSD as
\[
\Delta r^2_{\mathrm{tot}}(\Delta t)=
\left\langle \frac{1}{N}\sum_{i=1}^{N}
\left|\vec r_i(t+\Delta t)-\vec r_i(t)\right|^2
\right\rangle_t,
\]
where \(\langle \cdots \rangle_t\) denotes averaging over time origins. The total MSD can be decomposed as
\[
\langle \Delta r^2_{\mathrm{tot}}(\Delta t)\rangle
=
\langle \Delta r^2_{\mathrm{COM}}(\Delta t)\rangle
+
\langle \Delta r^2_{\mathrm{int}}(\Delta t)\rangle.
\]

In the motility-assay geometry, motors drive motion locally along the polymer contour. For a closed ring, these active forces do not generate a persistent net translational force on the polymer, so the COM motion remains predominantly diffusive (see Supplementary Fig. S1). The nontrivial activity dependence in \(\Delta r^2_{\mathrm{tot}}\) therefore arises primarily from internal modes and rotational motion of the ring. 

We first examine a semiflexible ring with persistence ratio \(l_p/L=0.5\). Figure~\ref{polyRing_on_MP_MSD_VaryPersistence}(a) shows \(\langle \Delta r^2_{\mathrm{tot}}(\Delta t)\rangle\) as a function of lag time \(\Delta t/\tau\) for different P\'eclet numbers \(Pe\). At long times (\(\Delta t/\tau \gtrsim 10^{-2}\)), the MSD is diffusive for all \(Pe\). At low activity (\(Pe=2.5\times 10^4\)), the short-time MSD exhibits subdiffusive growth, \(\Delta r^2_{\mathrm{tot}}\sim \Delta t^{3/4}\), followed by crossover to diffusion. Weak oscillatory modulations appear at later times, indicating the onset of rotational dynamics. At intermediate and high activity (\(Pe=6\times 10^5\) and \(1.5\times 10^6\)), the same short-time semiflexible scaling is observed over a progressively narrower window, followed by an intermediate ballistic regime, \(\Delta r^2_{\mathrm{tot}}\sim \Delta t^2\), and then diffusive growth at long times. The observed exponents and crossover trends are broadly consistent with theoretical expectations for active polymer rings~\cite{philipps2022dynamics}.
The oscillatory modulation becomes more pronounced with increasing activity, consistent with coherent ring rotation.

For a more flexible ring (\(l_p/L=0.03\)), the qualitative crossover structure remains similar, but the short-time behavior changes substantially [Fig.~\ref{polyRing_on_MP_MSD_VaryPersistence}(b)]. The \(\Delta t^{3/4}\) regime characteristic of semiflexible dynamics is no longer resolved, and the early-time MSD grows more slowly (\(\sim \Delta t^{0.6}\)). Upon systematically decreasing the persistence ratio (not shown), we find that this exponent decreases continuously and approaches \(1/2\) in the flexible limit, consistent with Rouse-like behavior~\cite{doi1988theory}. We also find that the MSD is only weakly affected by the bare processivity \(\Omega\) (see Supplementary Fig. S2), indicating that processivity primarily alters the effective strength and coherence of active forcing rather than the qualitative translational scaling structure. This weak processivity dependence of the MSD should be contrasted with the stronger \(\Omega\)-dependence of the rotational decorrelation time and asphericity discussed below. This contrast shows that processivity primarily modifies the temporal organization and intermittency of motor-mediated forcing rather than the gross translational scaling.


The MSD crossovers essentially provide a reference against which the motor-bed-specific features can be identified. These include the separation between rotational rate and rotational coherence, the non-monotonic decorrelation time at high processivity, and the activity-dependent reweighting of shape modes. These effects arise from stochastic motor binding, load-dependent stepping, and force-assisted detachment, and are analyzed in detail below.

\subsubsection{Autocorrelation of Ring Diameter: Signatures of Coherent Rotation}

\begin{figure*}[h!tbp]
\includegraphics[width=\textwidth]{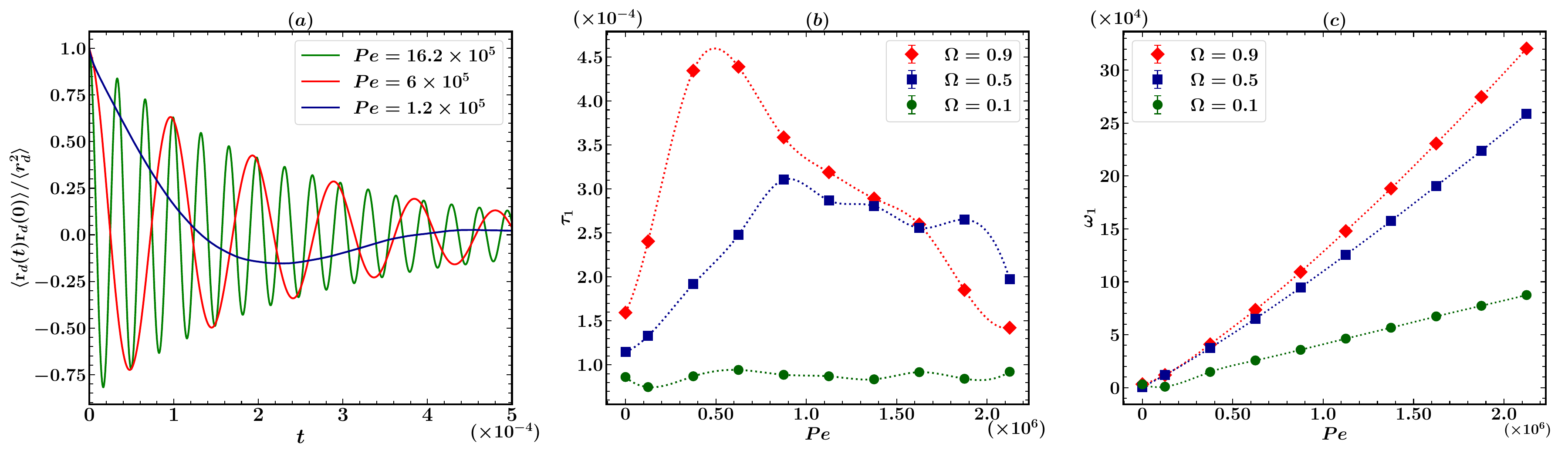}
\caption{\justifying 
The polymer persistence ratio is set to \(l_p/L = 0.5\). 
(a) Temporal autocorrelation of the ring diameter \(r_d(t)\) is shown for four different values of activity: \(Pe = 1.2\times10^5\), \(6\times10^5\), and \(16.2\times10^5\), keeping the bare processivity fixed at \(\Omega = 0.5\). The time \(t\) is expressed in units of \(\tau\).
(b) The dimensionless correlation decay timescale \(\tau_1\) is shown as a function of \(Pe\) for three different values of bare processivity: \(\Omega = 0.1\), \(0.5\), and \(0.9\). The dotted lines represent cubic spline interpolations and are shown only as guides to the eye.
(c) The dimensionless oscillation frequency \(\omega_1\) of the correlation function is plotted as a function of \(Pe\) for the same values of \(\Omega\).}
\label{polyRing_on_MP_radial_correlation}
\end{figure*}

To characterize the rotational dynamics underlying the oscillatory MSD features, we compute the autocorrelation of the ring-diameter vector \(\mathbf r_d(t)=\mathbf r(L/2,t)-\mathbf r(0,t)\), which connects two diametrically opposite points on the contour and captures the orientational memory of the ring as well as large-scale shape dynamics~\cite{mousavi2019active}. We define the normalized autocorrelation function as
\[
\mathcal C(t)=
\frac{\langle \mathbf r_d(t)\cdot \mathbf r_d(0)\rangle}
{\langle \mathbf r_d^2(0)\rangle}.
\]

Figure~\ref{polyRing_on_MP_radial_correlation}(a) shows \(\mathcal C(t)\) for a semiflexible ring (\(l_p/L=0.5\)) at fixed processivity and increasing activity. At low activity (\(Pe=1.2\times 10^5\)), \(\mathcal C(t)\) decays rapidly without a clear oscillatory component, indicating fast loss of orientational memory. As \(Pe\) increases (\(1.2\times 10^5\), \(6\times 10^5\), \(11.8\times 10^5\), \(16.2\times 10^5\)), damped oscillations emerge and the oscillation frequency increases systematically. This behavior reflects coherent rotational motion of the ring, with finite decorrelation due to thermal fluctuations, shape fluctuations, and stochastic motor kinetics. 
The appearance of oscillations itself is expected for tangentially driven active rings; the central question here is how explicit motor kinetics modifies the persistence and coherence of this rotation.
Across the range of processivities studied (\(\Omega=0.1,0.5,0.9\)), the diameter autocorrelation is well described by \(\mathcal C(t)\sim e^{-t/\tau_1}\cos(\omega_1 t)\),
where \(\tau_1\) is the decorrelation time and \(\omega_1\) is the oscillation frequency. We fit the measured \(\mathcal C(t)\) to this form at each \(Pe\) to extract \(\tau_1\) and \(\omega_1\). Figures~\ref{polyRing_on_MP_radial_correlation}(b) and (c) show the resulting \(Pe\)-dependence for fixed \(l_p/L=0.5\) and varying \(\Omega\).

\begin{figure*}[h!tbp]
\includegraphics[width=\textwidth]{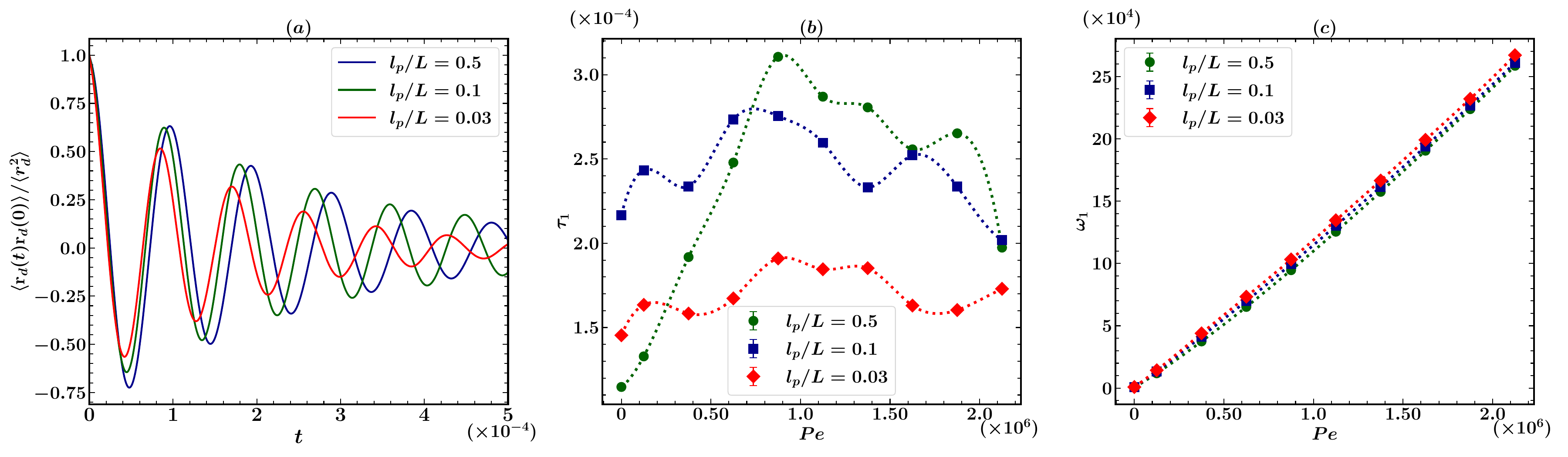}
\caption{\justifying 
The polymer bare processivity rate is set to \(\Omega = 0.5\). 
(a) Temporal autocorrelation of the ring diameter \(r_d(t)\) is shown for three different values of persistence ratio: \(l_p/L=0.5,0.1,0.03\), keeping activity \(Pe=6\times10^5\) fixed. The time \(t\) is expressed in units of \(\tau\).
(b) The dimensionless correlation decay timescale \(\tau_1\) is shown as a function of \(Pe\) for three different values of persistence ratio: \(l_p/L=0.5,0.1,0.03\). The dotted lines represent cubic spline interpolations and are shown only as guides to the eye.
(c) The dimensionless oscillation frequency \(\omega_1\) of the correlation function is plotted as a function of \(Pe\) for the same values of \(l_p/L\).}
\label{polyRing_on_MP_radial_correlation_varyPersistence}
\end{figure*}

The decorrelation time \(\tau_1\) exhibits a strong processivity dependence. For low processivity (\(\Omega=0.1\)), \(\tau_1\) varies weakly with activity. At intermediate and high processivity (\(\Omega=0.5\) and \(0.9\)), \(\tau_1\) becomes clearly non-monotonic in \(Pe\), with a peak at intermediate activity followed by a decrease at larger \(Pe\). This non-monotonicity is one of the main signatures of the motor-bed model. It shows that increasing activity does not simply make the ring rotate faster; it also changes how long the rotation remains coherent. We interpret this as a consequence of two competing effects: 
\((i)\) activity-induced synchronization of tangential motor forces, which stabilizes the sign of the net torque and enhances rotational coherence at intermediate \(Pe\), and 
\((ii)\) load-dependent stalling and detachment which are expected to generate stronger torque fluctuations and accelerate decorrelation at large \(Pe\). 

More explicitly, at low \(Pe\), the motor drive is weak, leading to weak and randomized rotation and hence a small \(\tau_1\).  
At intermediate \(Pe\), motor stepping is faster and generates a persistent tangential drive along the contour, yet remains sufficiently continuous to avoid strong intermittency. In a closed ring, a substantial fraction of bound motors can then experience comparable tangential loads, promoting synchronized motion along the contour. Because the ring is closed, motors step in a fixed contour direction while attached, so the instantaneous net torque can maintain its sign over longer times. Consequently, the diameter vector retains its direction over multiple rotation periods. In this regime, increasing activity enhances rotational coherence.

At large \(Pe\), the dynamics become more intermittent. Since both stepping velocity and detachment are load-dependent, strong elastic forces at high \(Pe\) can cause some motors to stall (large \(f_t\)) while others detach (large \(f_l\)). This produces a heterogeneous, time-dependent distribution of effective drive along the filament. For a linear filament, stress can be relieved through end motion or curvature redistribution. In a ring, however, the absence of free ends couples local stress buildup to global shape fluctuations. These shape fluctuations, in turn, modify local tangents, motor loads, and torque generation. Although the ring may still rotate rapidly (large \(\omega_1\)), fluctuations in the instantaneous torque lead to faster decorrelation of the rotation axis and diameter vector, resulting in a reduced \(\tau_1\).

The oscillation frequency \(\omega_1\) increases monotonically with activity for all processivities, with a steeper increase at larger \(\Omega\) (see Fig.~\ref{polyRing_on_MP_radial_correlation}(c)). 
Thus, the motor bed separates rotational speed from rotational coherence: \(\omega_1\) continues to increase with activity, while \(\tau_1\) decreases once load-induced intermittency and shape fluctuations dominate. This behavior contrasts with idealized active-polar-ring models with prescribed tangential self-propulsion. There the rotational dynamics is controlled by a simpler activity-dependent frequency and the equal-time conformations are not modified in the continuum limit~\cite{philipps2022dynamics}. In the present motility-assay model, the stochastic binding/unbinding of motors and the load dependence of both stepping and detachment introduce additional timescales and force heterogeneities, producing a richer dependence on \(Pe\) and \(\Omega\).

We next examine the role of stiffness by computing \(\mathcal C(t)\) for persistence ratios spanning the semiflexible-to-flexible regime. As shown in Fig.~\ref{polyRing_on_MP_radial_correlation_varyPersistence}(a), softer rings exhibit faster decay of the oscillatory envelope, indicating more rapid loss of orientational memory due to enhanced shape fluctuations. Consistent with this, Fig.~\ref{polyRing_on_MP_radial_correlation_varyPersistence}(b) shows that the non-monotonic dependence of \(\tau_1\) on \(Pe\) is strongest for stiffer rings and becomes weaker as the ring becomes more flexible. Flexible rings therefore display more weakly coherent rotational dynamics, with decorrelation dominated by internal fluctuations. This further supports the interpretation that \(\tau_1\) is controlled not only by the average motor-driven rotation rate, but also by the magnitude of shape fluctuations that randomize the orientation of the diameter vector.

Fig.~\ref{polyRing_on_MP_radial_correlation_varyPersistence}(c) shows that \(\omega_1\) depends only weakly on the persistence ratio over the range explored. This suggests that the characteristic rotational rate is controlled primarily by motor-driven forcing (via \(Pe\) and \(\Omega\)), whereas polymer stiffness predominantly modulates the coherence of the rotation rather than its mean frequency. 
This separation is consistent with the ring-size analysis below: over the stiffness range studied, the typical ring size changes only modestly compared with the change in the damping of the autocorrelation. Thus, stiffness mainly affects the fluctuation-induced loss of phase coherence rather than the average rotational drift.
Stiffness therefore mainly renormalizes the damping of the oscillation, while the oscillation frequency reflects the average motor-induced rotational drift of the ring’s diameter axis. 
Having established this separation of roles, we next examine an independent kinematic observable, the instantaneous angular velocity in the center-of-mass frame, to corroborate the rotational trends inferred from the diameter autocorrelation.

\subsubsection{Angular Velocity of Polymer Ring}

As an independent kinematic measure of rotation, we compute the instantaneous angular velocity of the ring in the center-of-mass (COM) frame from monomer positions and velocities. For each monomer, we define \((x_i,y_i)\) and \((v_{x,i},v_{y,i})\) relative to the instantaneous COM position and COM velocity, and evaluate the local angular rate
\[
\omega_i(t)=\frac{x_i v_{y,i}-y_i v_{x,i}}{x_i^2+y_i^2}.
\]
We then define the net angular velocity as
\[
\omega_{\mathrm{ang}}(t)=\frac{1}{N}\sum_{i=1}^{N}\omega_i(t),
\]
and analyze its time-averaged magnitude, \(\langle |\omega_{\mathrm{ang}}| \rangle\), as a function of activity and motor processivity. This observable probes the same underlying rotational dynamics as the oscillation frequency \(\omega_1\) extracted from the diameter autocorrelation, but provides a direct kinematic diagnostic rather than a correlation-based one.

\begin{figure}[htb!]
\includegraphics[width=\columnwidth]{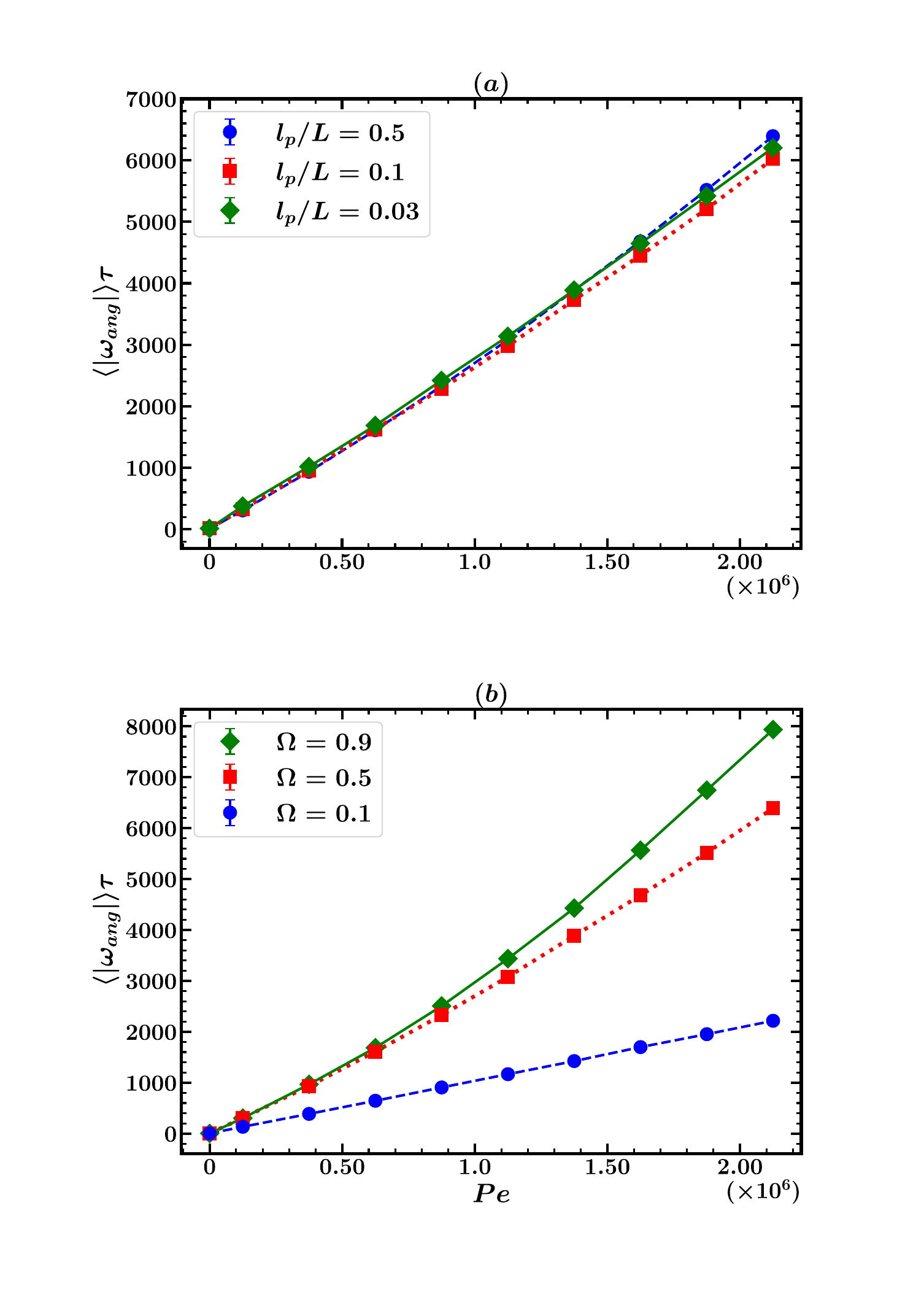}
\caption{\justifying  (a) Net dimensionless angular velocity $\langle|\omega_{\rm ang}|\rangle \tau$ as a function of activity $Pe$ for different persistence ratios $l_p/L$ at fixed motor processivity $\Omega=0.5$. (b) Net angular velocity $\langle|\omega_{\rm ang}|\rangle$ versus $Pe$ for three motor processivity rates $\Omega = 0.1, 0.5, 0.9$ at fixed persistence ratio $l_p/L=0.5$.}
\label{polyRing_on_MP_v=24_angularVel}
\end{figure}

Figure~\ref{polyRing_on_MP_v=24_angularVel}(a) shows \(\langle |\omega_{\mathrm{ang}}| \rangle\) versus \(Pe\) for several persistence ratios at fixed processivity \(\Omega=0.5\). The average angular velocity increases monotonically with activity, confirming that stronger motor-driven forcing enhances ring rotation. The dependence on persistence ratio is weak over the range studied, consistent with the diameter-autocorrelation analysis, where the rotational frequency \(\omega_1\) was also found to be only weakly stiffness dependent. Together, these results indicate that stiffness primarily controls the damping and persistence of rotational motion (through decorrelation), rather than the characteristic rotation rate.

The angular-velocity analysis also reinforces the distinction between rotational speed and coherence. The monotonic increase of \(\langle |\omega_{\mathrm{ang}}| \rangle\) with \(Pe\) shows that the motors continue to drive faster contour-level rotation at large activity. However, the simultaneous decrease of \(\tau_1\) at high \(Pe\) and high \(\Omega\) demonstrates that this faster rotation becomes less coherent because of load-induced intermittency and enhanced shape fluctuations.

The effect of motor processivity is shown in Fig.~\ref{polyRing_on_MP_v=24_angularVel}(b) for a fixed persistence ratio \(l_p/L=0.5\). At low processivity, \(\langle |\omega_{\mathrm{ang}}| \rangle\) increases approximately linearly with \(Pe\), whereas at higher processivity the growth becomes nonlinear, with stronger enhancement at large activity. This trend is consistent with the processivity dependence inferred from the diameter autocorrelation: increasing \(\Omega\) not only raises the rotational rate but also changes the temporal organization of motor-driven motion around the closed contour.

\begin{figure*}[h!tbp]
\includegraphics[width=0.9\textwidth]{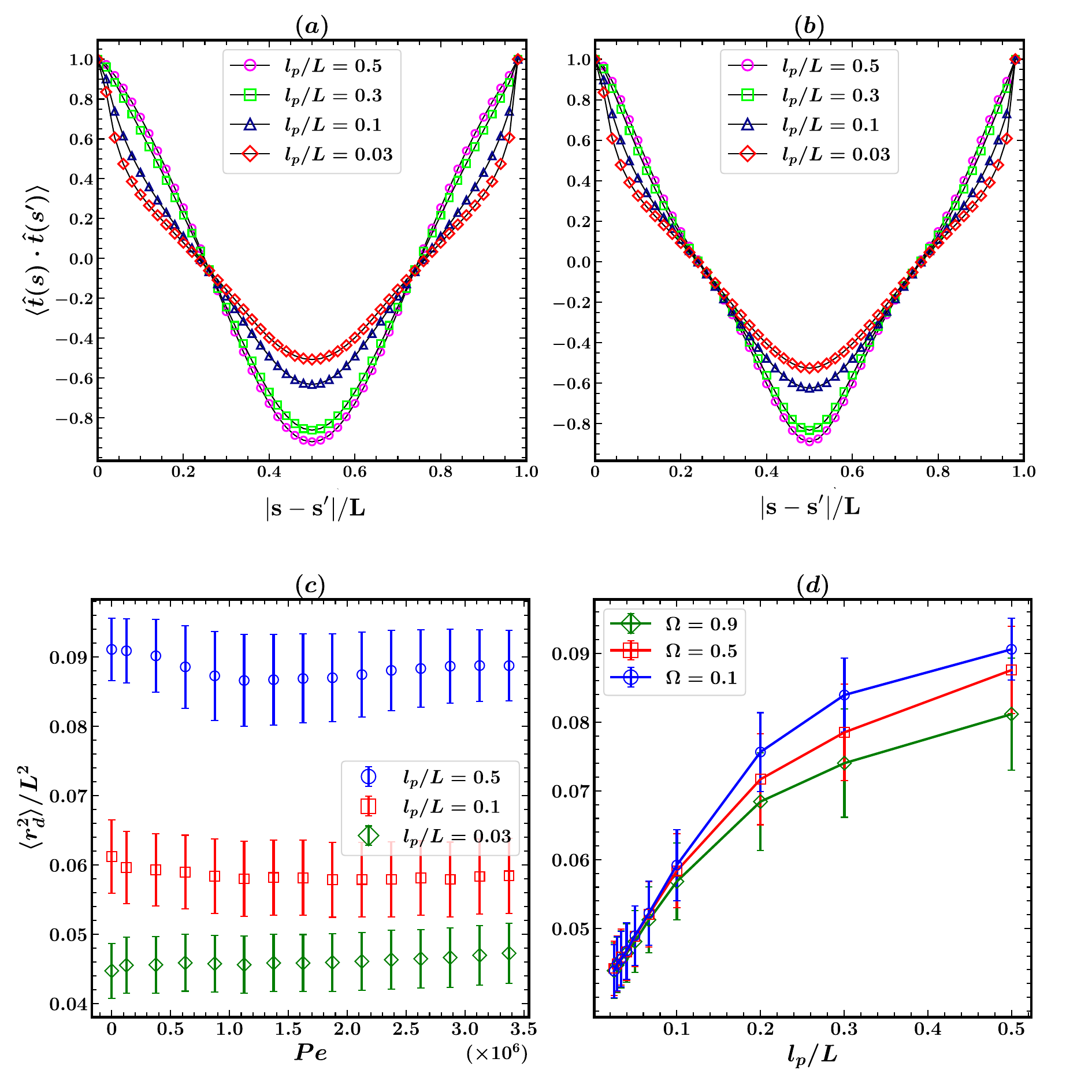}
\caption{\justifying
Tangent--tangent spatial correlation of a semiflexible ring polymer is shown. Four persistence ratios are considered: $l_p/L = 0.5$, $0.3$, $0.1$, and $0.03$. Panels (a) and (b) show the correlation for fixed activity $Pe = 6 \times 10^5$ at two bare processivity rates, $\Omega = 0.2$ and $0.8$, respectively. Panels (c) and (d) show the normalized mean squared diameter, $\langle r_d^2 \rangle/L^2$, as a function of $Pe$ and $l_p/L$ . In (c), the persistence ratio is varied ($l_p/L = 0.03$, $0.1$, $0.5$) with $\Omega = 0.5$, while in (d), $\Omega$ is varied ($0.1$, $0.5$, $0.9$) with fixed $Pe = 8.75\times10^5$.}
\label{polyRing_on_MP_tantan_corr}
\end{figure*}

\begin{figure*}[h!tbp]
\includegraphics[width=\textwidth]{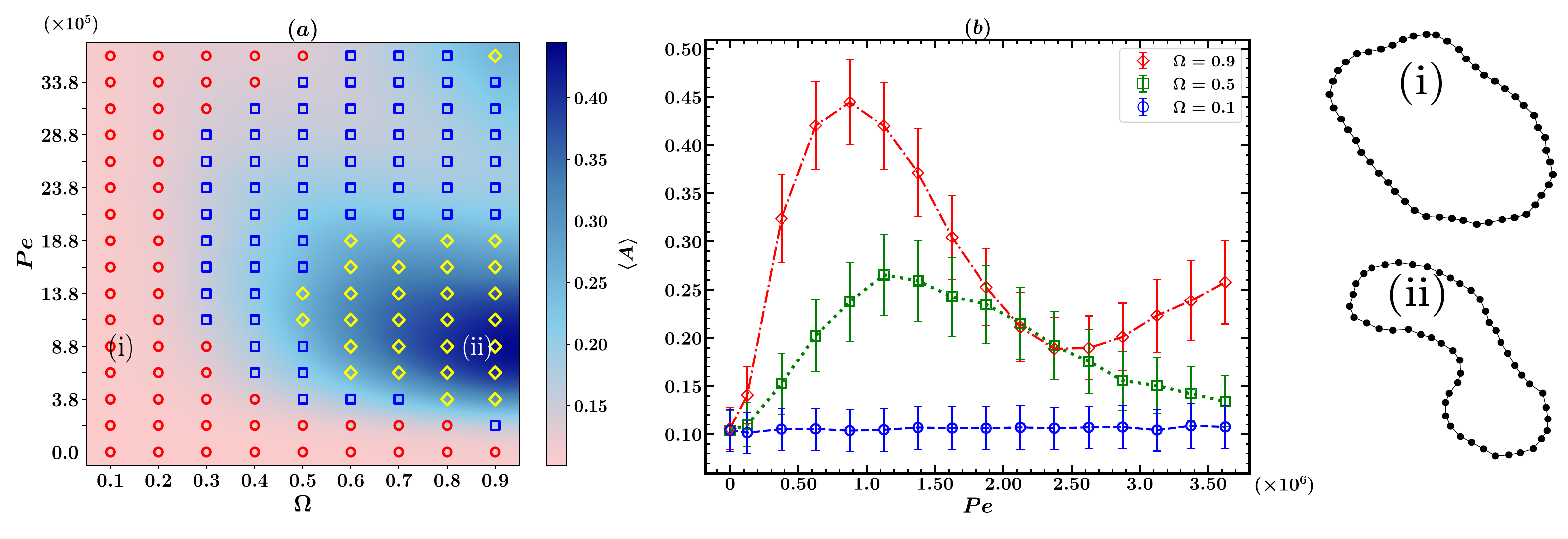}
\caption{\justifying
Average asphericity $\langle A \rangle$ of a semiflexible ring polymer of persistence ratio $l_p/L = 0.5$, is plotted. 
(a) Phase diagram of $\langle A \rangle$ as a function of both $Pe$ and $\Omega$, with the persistence ratio fixed at $l_p/L = 0.5$.
(b) Variation of $\langle A \rangle$ as a function of the Peclet number ($Pe$) for three different values of the bare processivity rate: $\Omega = 0.1$, $0.5$, and $0.9$. 
}
\label{polyRing_on_MP_asphericity_phaseDiag}
\end{figure*}

\begin{figure*}[h!tbp]
\includegraphics[width=\textwidth]{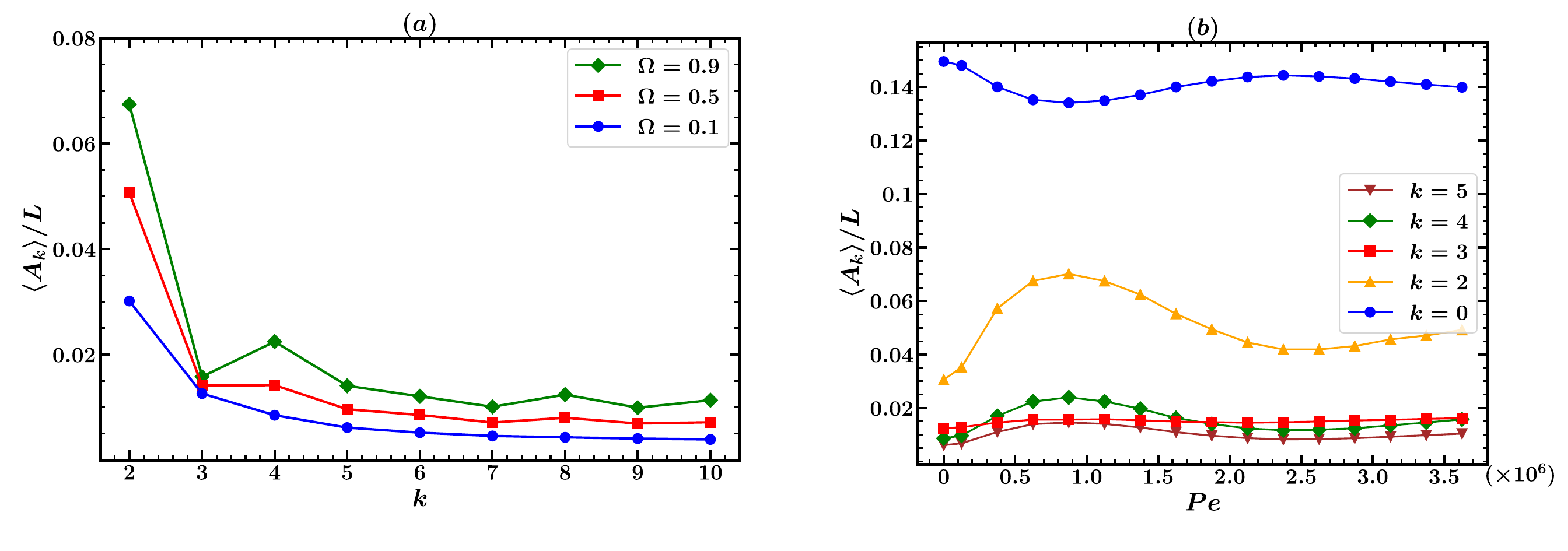}
\caption{\justifying Average mode amplitude, $\langle A_k \rangle/L$, of a semiflexible ring polymer with persistence ratio $l_p/L = 0.5$.
(a) Dependence of $\langle A_k \rangle/L$ on the mode number $k$ for different processivity rates $\Omega = 0.1,0.5,0.9$, at fixed Peclet number $Pe = 11.25 \times 10^5$.
(b) Dependence of $\langle A_k \rangle/L$ on the Peclet number $Pe$ for selected modes ($k=0,2,3,4,5$), at fixed processivity rate $\Omega = 0.9$.}
\label{fig:modes}
\end{figure*}

\subsection{Shape Analysis}
\label{sec:Shape}

\subsubsection{Tangent--Tangent Correlation}

Having discussed the transport and rotational dynamics, we next examine the conformational statistics of the ring. As a first measure, we compute the tangent--tangent correlation function, $C_t(\Delta s) = \langle \hat{\mathbf t}(s)\cdot \hat{\mathbf t}(s')\rangle$, where \(\hat{\mathbf t}(s)\) is the unit tangent vector at contour position \(s\) and $\Delta s = s - s^\prime$. This correlation probes orientational memory along the contour and is sensitive to both polymer stiffness and the symmetry of ring conformations.

Figure~\ref{polyRing_on_MP_tantan_corr}(a,b) shows the tangent--tangent correlation as a function of arc-length separation \(|s-s'|\) for persistence ratios \(l_p/L=0.03,\,0.1,\,0.3,\) and \(0.5\), at fixed activity \(Pe=6\times 10^5\) and two processivities, \(\Omega=0.2\) and \(0.8\). For \(\Omega=0.2\) (see Fig.~\ref{polyRing_on_MP_tantan_corr}(a)), the correlation profiles are symmetric about \(|s-s'|=L/2\), as expected for a closed ring. The curves exhibit a smooth well-like structure, with two symmetric intersections across stiffness values. As the persistence ratio decreases (more flexible rings), the correlation well broadens and the minimum becomes shallower, reflecting enhanced conformational fluctuations. By contrast, stiffer rings show slower initial decay but a sharper drop near \(|s-s'|\simeq L/2\), producing a deeper minimum associated with stronger contour-scale orientational organization.

At higher processivity, \(\Omega=0.8\) (see Fig.~\ref{polyRing_on_MP_tantan_corr}(b)), the same stiffness-dependent trends persist, but the intersection points broaden into wider regions of approximate overlap. The overall profiles become less sharply differentiated at intermediate arc-length separations, consistent with stronger motor-driven contour reorganization. The influence of stiffness, however, remains clear: stiffer rings retain more structured correlation profiles with deeper minima, whereas flexible rings exhibit smoother, more weakly structured decay.

 For an ideal smooth circular ring, the tangent correlation reduces to a cosine-like form, \(C_t(\Delta s)\simeq \cos(2\pi \Delta s/L)\), and the \(l_p/L=0.5\) curve is close to this stiff-ring limit. The sharper endpoint variation observed for more flexible rings is due to the growing contribution of higher Fourier modes to the tangent correlation. This is consistent with the analytical active-polar-ring theory of Philipps et al~\cite{philipps2022dynamics}. In the present bead--spring model, the correlation is evaluated using discrete bond tangents. Therefore, the endpoint behavior appears as a steep initial decay over the first few bond separations rather than the smooth continuum limit.

Together, these results show that stiffness primarily controls the contour-scale orientational structure of the ring, while motor processivity modulates how sharply these conformational correlations are maintained under active driving.

\subsubsection{Activity-Induced Swelling and Shrinkage of the Polymer}

To quantify changes in ring size and compaction, we compute the normalized mean squared diameter, \(\langle r_d^2\rangle/L^2\), as a function of activity \(Pe\). This quantity provides a direct measure of swelling or shrinkage of the ring under motor-driven forcing.

Figure~\ref{polyRing_on_MP_tantan_corr}(c) shows \(\langle r_d^2\rangle/L^2\) versus \(Pe\) at fixed processivity \(\Omega=0.5\) for several persistence ratios. The average diameter decreases as the ring becomes more flexible (smaller \(l_p/L\)), indicating stronger compaction in softer polymers. For the most flexible cases (\(l_p/L=0.03\) and \(0.1\)), \(\langle r_d^2\rangle/L^2\) varies only weakly with activity, suggesting that large shape fluctuations already dominate the ring size and suppress further activity-induced changes. 
In contrast, for the stiffest ring ($l_p/L=0.5$), $\langle r_d^2\rangle/L^2$ remains nearly constant at low $Pe$, shows a slight reduction at intermediate $Pe$, and increases again at larger $Pe$. However, the intermediate dip lies within statistical uncertainty, 
so the overall size variation is weak and consistent with an approximately constant response. Overall, however, the magnitude of the size variation remains modest. 

The absence of a strong activity-induced collapse in our system while similar to that in Ref.~\cite{philipps2022dynamics}, is in contrast to the behavior reported in Ref.~\cite{locatelli2021activity}. Our system differs in several respects: the ring is semiflexible, short ($N = 50$), confined to a two-dimensional motility-assay geometry, and driven by stochastic motor attachment, stepping, and detachment rather than by a prescribed normalized tangential force on every monomer.

Figure~\ref{polyRing_on_MP_tantan_corr}(d) shows \(\langle r_d^2\rangle/L^2\) as a function of persistence ratio \(l_p/L\) at fixed activity \(Pe=8.75\times 10^5\), for three values of the bare processivity \(\Omega\). As stiffness increases (larger \(l_p/L\)), the normalized diameter rises rapidly at first and then approaches a weaker dependence, indicating a gradual saturation of ring size in the stiff-ring limit. At fixed stiffness, \(\langle r_d^2\rangle/L^2\) decreases systematically with increasing \(\Omega\). Thus, higher processivity yields a slightly smaller mean diameter, with the separation between the curves becoming more pronounced for stiffer rings. With high processivity, more motors remain bound for longer. This leads to sustained forcing along the contour, weakly promoting additional compaction of the ring.

\subsubsection{Asphericity}

To quantify deviations from circular shape, we compute the asphericity from the radius-of-gyration tensor,
\[
G_{\alpha\beta}=\frac{1}{N}\sum_{i=1}^{N}
\big(r_{i,\alpha}-r_{\mathrm{cm},\alpha}\big)
\big(r_{i,\beta}-r_{\mathrm{cm},\beta}\big),
\]
where \(r_{i,\alpha}\) is the \(\alpha\)-component of monomer \(i\), and \(r_{\mathrm{cm},\alpha}\) is the corresponding center-of-mass component. In two dimensions, \(G\) has eigenvalues \(\lambda_1\) and \(\lambda_2\), and the asphericity is defined as~\cite{rudnick1986aspherity}
\[
A=\frac{(\lambda_1-\lambda_2)^2}{(\lambda_1+\lambda_2)^2}.
\]
A perfectly circular ring has \(A=0\), while elongated or distorted conformations yield larger \(A\).

We map the average asphericity \(\langle A\rangle\) as a function of activity \(Pe\) and bare processivity \(\Omega\) in Fig.~\ref{polyRing_on_MP_asphericity_phaseDiag}(a). The symbols/colors indicate three ranges of \(\langle A\rangle\): nearly circular (\(\langle A\rangle<0.14\)), intermediate distortion (\(0.14<\langle A\rangle<0.25\)), and strongly distorted (\(\langle A\rangle>0.25\)). Although the threshold values are chosen empirically, they provide a convenient way to visualize the crossover between shape regimes. Representative conformations included in the figure illustrate the corresponding nearly circular and elongated ring states.

At low processivity (\(\Omega\lesssim 0.1\)), \(\langle A\rangle\) remains small over the full activity range, indicating weak shape distortion (see Supplementary Movie I). However, at higher processivity, \(\langle A\rangle\) shows a pronounced non-monotonic dependence on \(Pe\), as seen more clearly in Fig.~\ref{polyRing_on_MP_asphericity_phaseDiag}(b) (see Supplementary Movie II and see Supplementary Fig. S3 for length dependence). For \(\Omega=0.5\) and \(0.9\), \(\langle A\rangle\) increases at low activity, reaches a maximum, decreases at intermediate \(Pe\), and rises again at larger \(Pe\). The position of the maximum shifts to lower \(Pe\) as \(\Omega\) increases, indicating that stronger motor processivity promotes maximal shape deformation at lower activity.

To rationalize the non-monotonic behavior of \(\langle A\rangle\), we analyze ring conformations in the center-of-mass (COM) frame using a Fourier decomposition in polar coordinates. Writing the bead positions as \((r_i,\theta_i)\), we define
\[
a_k=\frac{1}{N}\sum_{i=1}^{N} r_i\cos(k\theta_i),\qquad
b_k=\frac{1}{N}\sum_{i=1}^{N} r_i\sin(k\theta_i),
\]
and the corresponding mode amplitude
\[
A_k=\sqrt{a_k^2+b_k^2}.
\]
Here, \(A_k\) has dimensions of length and quantifies the strength of the \(k\)-th shape mode. Fig.~\ref{fig:modes}(a) shows the average mode amplitude $\langle A_k \rangle$ as a function of mode number $k$ for three processivity values, $\Omega = 0.1, 0.5,$ and $0.9$. 
At large processivity, the lower modes exhibit enhanced amplitudes, with the $k=2$ mode, corresponding to elliptic distortions, being particularly dominant, while higher modes represent progressively more corrugated deformations. As processivity decreases, the contribution from these shape modes diminishes, and the $k=0$ mode (the mean ring radius) captures the overall configuration. This trend is consistent with the asphericity behavior shown in Fig.~\ref{polyRing_on_MP_asphericity_phaseDiag}(a).

The dependence of individual mode amplitudes on the P\'eclet number is shown in Fig.~\ref{fig:modes}(b) for $\Omega=0.9$. The non-monotonic variation in asphericity is reflected primarily in the $k=2$ mode amplitude, which peaks at intermediate P\'eclet numbers. In contrast, the $k=0$ mode exhibits a slight dip in the same $Pe$ range. Overall, the mode analysis indicates that elliptic distortions ($k=2$) constitute the dominant rotational deformation at intermediate P\'eclet numbers and large processivity, where bound motors are distributed relatively homogeneously along the contour. 
This behavior, however, depends on the persistence ratio, which is fixed here at $l_p/L=0.5$.

\begin{figure}[h!]
\centering
\includegraphics[width=\columnwidth]{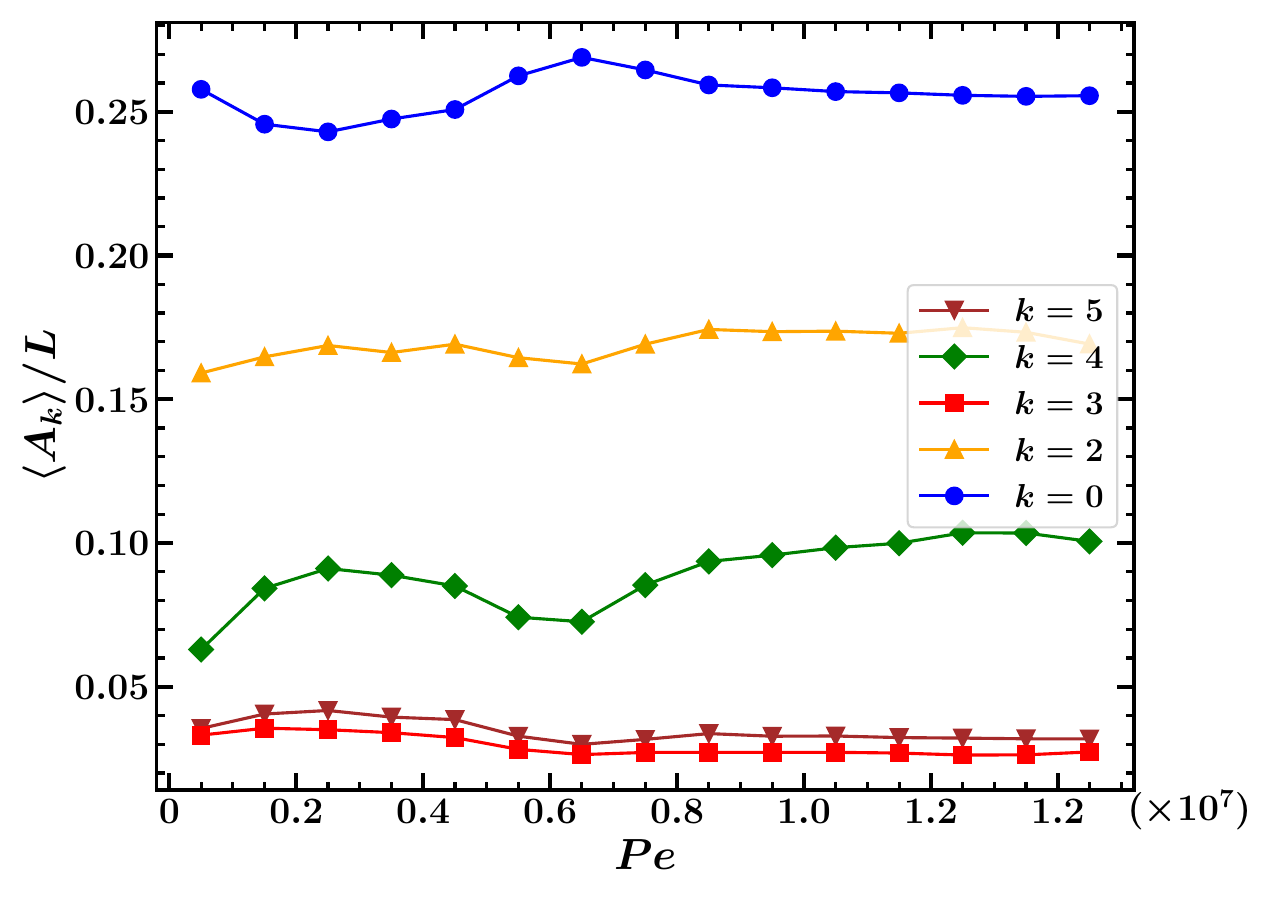}
\caption{\justifying The average mode amplitude \(\langle A_k \rangle/L\) as a function of the P\'eclet number \(Pe\), with the bare processivity rate fixed at \(\Omega = 0.9\) and persistence ratio \(l_p/L = 0.5\). The polymer length is fixed at \(L = 100\sigma\).}
\label{fig:asphericity_pe_varyL100}
\end{figure}

To test whether the \(k=2\)-dominated response is specific to the short ring, we repeated the mode analysis for a longer ring with \(L=100\sigma\) (Fig.~\ref{fig:asphericity_pe_varyL100}). We observe that the magnitude of the average mode amplitudes is larger than that for the shorter ring. The behavior of the \(k=0\) mode remains qualitatively similar to the \(L=50\sigma\) case. However, in contrast to the \(k=2\) mode for \(L=50\sigma\), where a pronounced nonmonotonic dependence on \(Pe\) was observed, the nonmonotonicity for \(k=2\) at \(L=100\sigma\) is strongly suppressed, and the mode amplitude remains nearly constant over the entire range of \(Pe\). Interestingly, the \(k=4\) mode exhibits a more prominent nonmonotonic behavior in the longer polymer ring. This indicates that longer rings redistribute activity-induced deformation into higher-order modes.

\section{Summary and Outlook}
\label{sec:SummaryOutlook}

We have studied the nonequilibrium dynamics and conformations of a passive semiflexible polymer ring driven by a bed of active motor proteins, using a coarse-grained model that explicitly incorporates stochastic motor attachment--detachment kinetics and load-dependent stepping. This framework complements recent studies of intrinsically active polymer rings~\cite{philipps2022dynamics,locatelli2021activity} by introducing motor-mediated forcing and processivity as independent control parameters. The model should be viewed primarily as a minimal two-dimensional motility-assay realization of a closed motor-driven filament, rather than as a direct model of intracellular ring-like polymers. In this sense, it provides a controlled setting for isolating how closed topology, filament elasticity, and stochastic motor kinetics combine to shape nonequilibrium ring dynamics.

Our results show that motor-driven ring polymers exhibit a rich hierarchy of dynamical regimes governed by activity, stiffness, and motor processivity. The mean-squared displacement displays the expected crossover structure of active polymer motion, with subdiffusive-to-diffusive behavior at low activity and an intermediate ballistic regime emerging at higher \(Pe\). For more flexible rings, the short-time semiflexible scaling is replaced by slower growth that approaches the Rouse-like limit, while the ballistic regime remains visible at moderate activity. These trends are broadly consistent with theoretical expectations for active rings~\cite{philipps2022dynamics}. We therefore regard the MSD primarily as a baseline dynamical diagnostic, rather than as the central novelty of the work. The motility-assay setting adds a distinct rotational component: diameter autocorrelations develop damped oscillations whose frequency \(\omega_1\) increases monotonically with activity, whereas the decorrelation time \(\tau_1\) shows a pronounced processivity-dependent non-monotonicity. This separation between rotational rate and rotational coherence is a central feature of the motor-driven ring dynamics. A direct kinematic estimator based on the instantaneous angular velocity corroborates the same trend, showing weak stiffness dependence of the rotational rate but strong sensitivity to motor processivity.

The conformational response is likewise controlled by the interplay of stiffness and motor kinetics. Tangent--tangent correlations reveal how increasing flexibility smooths contour-scale orientational structure, while higher processivity broadens the correlation profiles through stronger motor-driven reorganization. Ring size, quantified by the normalized mean squared diameter, remains nearly unchanged with activity for flexible rings but shows a weak non-monotonic shrinkage--reswelling trend for stiffer rings. The most striking structural signature is the asphericity, which becomes strongly non-monotonic in \(Pe\) at high processivity. Fourier-mode analysis in the center-of-mass frame shows that this behavior arises from an activity-dependent reweighting of the ring-shape fluctuations among Fourier modes, in particular the competition between the mean-radius (\(k=0\)) and elliptic (\(k=2\)) modes, with increasing contributions from higher modes at large activity.

More broadly, our results highlight the qualitative impact of motor binding kinetics on active-polymer dynamics. In contrast to idealized active-ring models, where relaxation times often show simpler trends with activity~\cite{philipps2022dynamics}, the present system exhibits nontrivial and non-monotonic behavior because motor attachment, detachment, and load-dependent stepping introduce additional dynamical timescales. At intermediate activity, persistent motor stepping supports coherent contour-following rotation. At higher activity, load-induced stalling, force-assisted detachment, and heterogeneous motor loading amplify torque and shape fluctuations. This reduces rotational coherence even though the mean rotational speed continues to increase. This distinction between rotational speed and rotational coherence is one of the main achievements of this paper.

The biological motivation for studying closed active filaments should, however, be interpreted with caution. The present model does not include three-dimensional conformations, hydrodynamic interactions, background cytoskeletal crowding, topological threading, or entanglements with surrounding filaments. These ingredients can strongly affect the transport and conformations of circular DNA or ring polymers in active cytoskeletal networks. Thus, our results should not be read as a direct description of intracellular circular DNA, contractile rings, or cytoskeletal ring structures in their full cellular complexity. Instead, the present model is complementary to such systems: it isolates the simpler limit of a single closed semiflexible contour driven by stochastic, load-dependent motor interactions on a two-dimensional substrate.

A direct experimental realization of the present predictions would be most natural in reconstituted motility assays or engineered biomimetic motor--filament platforms. Ring- or loop-like actin filaments or microtubules moving on myosin- or kinesin-coated substrates would provide an appropriate setting in which motor density, ATP concentration, filament stiffness, and motor processivity could be varied systematically. Time-resolved imaging of such assays could provide access to the same observables analyzed here. In particular, the key prediction to test is that explicit motor kinetics can separate rotational speed from rotational coherence and generate non-monotonic shape responses beyond those expected from continuously driven active-ring models.

Finally, the phenomenology reported here connects naturally to a broader class of ring- and loop-like active soft-matter systems. Ring and loop formation in gliding assays has been reported for microtubule--motor systems~\cite{kawamura2008ring, liu2011loop}, and related collective organization principles also appear in synthetic biomolecular engines~\cite{keya2020synchronous}. While cellular structures such as marginal bands~\cite{dmitrieff2017balance} and actomyosin contractile rings during cytokinesis~\cite{pinto2012actin} involve additional biochemical, mechanical, and geometrical complexity, they provide broader motivation for understanding how activity and closed topology interact. In this sense, motor-driven polymer rings on motility assays provide a minimal but versatile platform for isolating how activity, filament elasticity, and binding kinetics combine to generate emergent rotational and shape dynamics in active soft matter.

\section*{Supplementary Material}
\begin{figure}[h!tbp]
\includegraphics[width=\columnwidth]{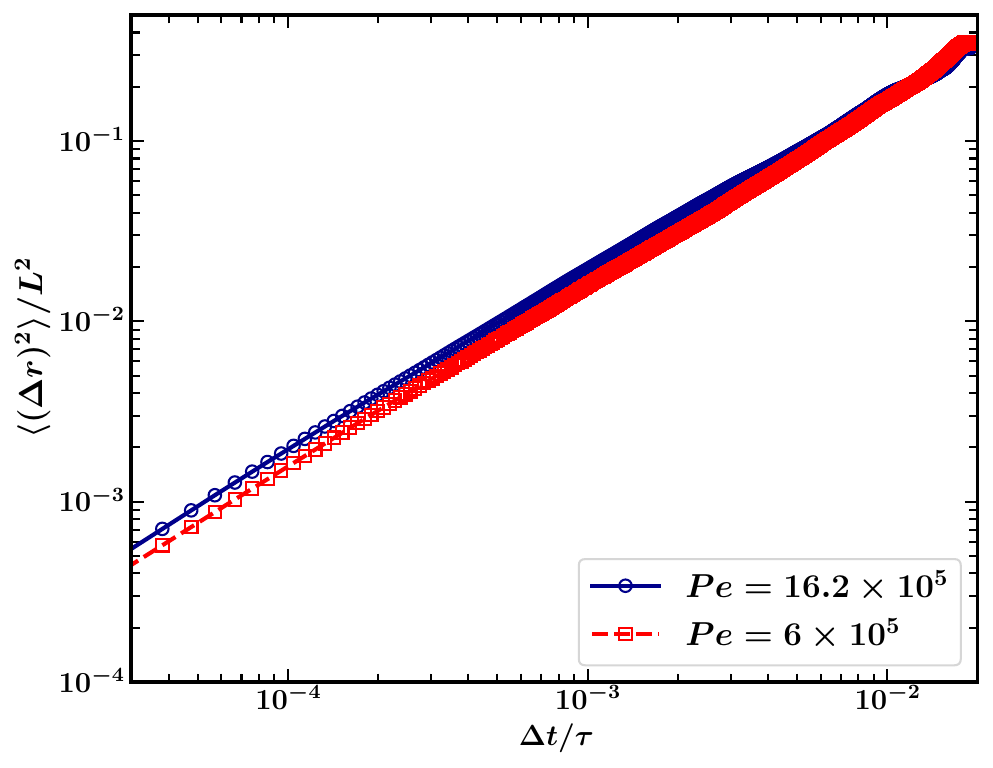}
\caption{\justifying Time evolution of the mean squared displacement (MSD) of the center of mass (COM) of a semiflexible polymer ring on a motility assay. The bare processivity rate is fixed at $\Omega=0.5$. Results are shown for two activity strengths corresponding to $Pe = 6\times10^5$ and $Pe = 16.2\times10^5$.}
\label{fig:COM_MSD}
\end{figure}

\section{Center-of-mass transport}
\label{app:COM}
In the main text we emphasized that, for a closed ring, contour-directed motor forces do not generate a persistent net translational force, and the activity dependence of the total MSD primarily reflects internal and rotational modes. To verify this explicitly, Fig.~\ref{fig:COM_MSD} shows the mean-squared displacement of the ring center of mass (COM) for two representative activity strengths at fixed processivity. The COM MSD remains close to diffusive scaling over the accessible time window, with only weak dependence on activity. This confirms that motor activity predominantly reorganizes internal conformations and rotational dynamics rather than producing sustained self-propulsion of the COM. Consequently, the oscillatory features and crossover regimes observed in the total MSD arise mainly from internal ring modes and coherent rotation.

\section{Mean-squared displacement at different processivities}

\begin{figure*}[!htb]
\centering
\includegraphics[width=\textwidth]{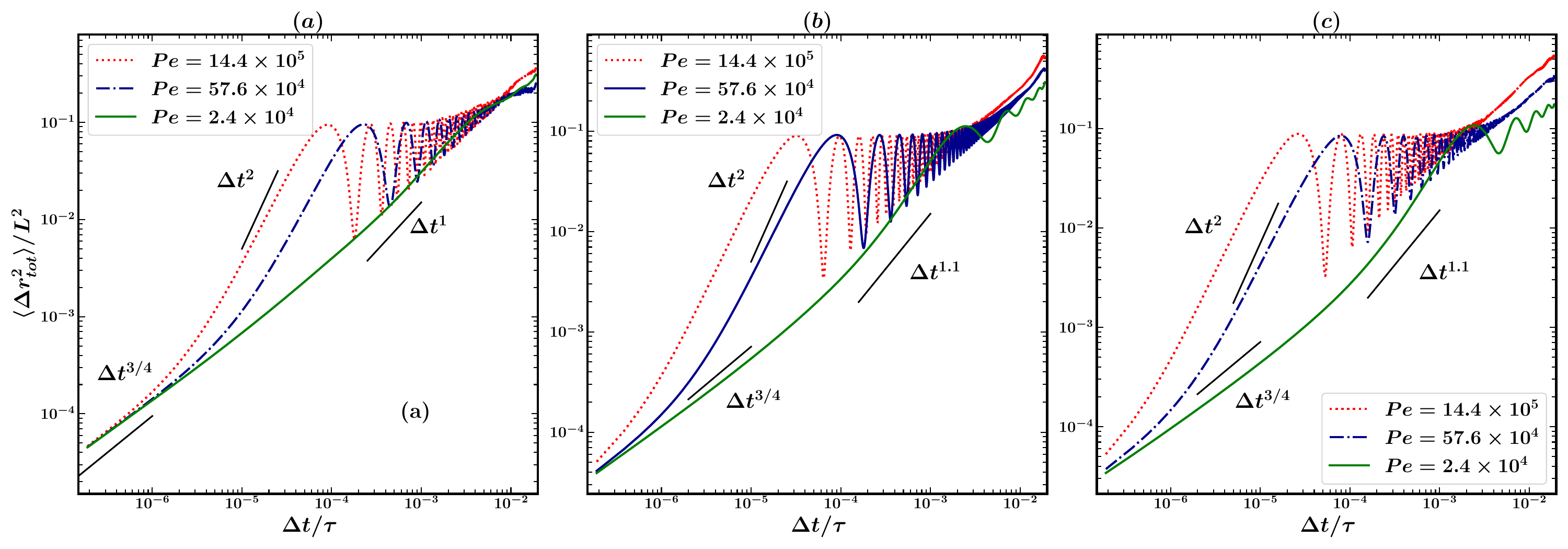}
\caption{\justifying Mean squared displacement (MSD) of a semiflexible polymer ring in a motility assay at a fixed persistence ratio of \(l_p/L = 0.5\). The three panels correspond to different values of the bare processivity rate $\Omega$: (a) $\Omega=0.1$, (b) $\Omega=0.5$, and (c) $\Omega=0.9$.}
\label{fig:MSD_vary_Omega}
\end{figure*}

 Figure~\ref{fig:MSD_vary_Omega} shows the total MSD at fixed stiffness for three values of the bare processivity, \(\Omega=0.1\), \(0.5\), and \(0.9\). Across all processivities, the MSD exhibits the same qualitative sequence of regimes: a short-time semiflexible subdiffusive scaling followed by an intermediate ballistic window at sufficiently large activity, and eventual long-time diffusion. The primary effect of increasing \(\Omega\) is to enhance the prominence of the intermediate-time active regime and the associated oscillatory signatures, consistent with stronger and more persistent motor engagement. Thus, the ring MSD is robust to changes in \(\Omega\). This is consistent with our claim that processivity essentially modulates the effective strength and temporal organization of active forcing.

\section{Dependence on ring length}
\begin{figure}[h!tbp]
\centering
\includegraphics[width=\columnwidth]{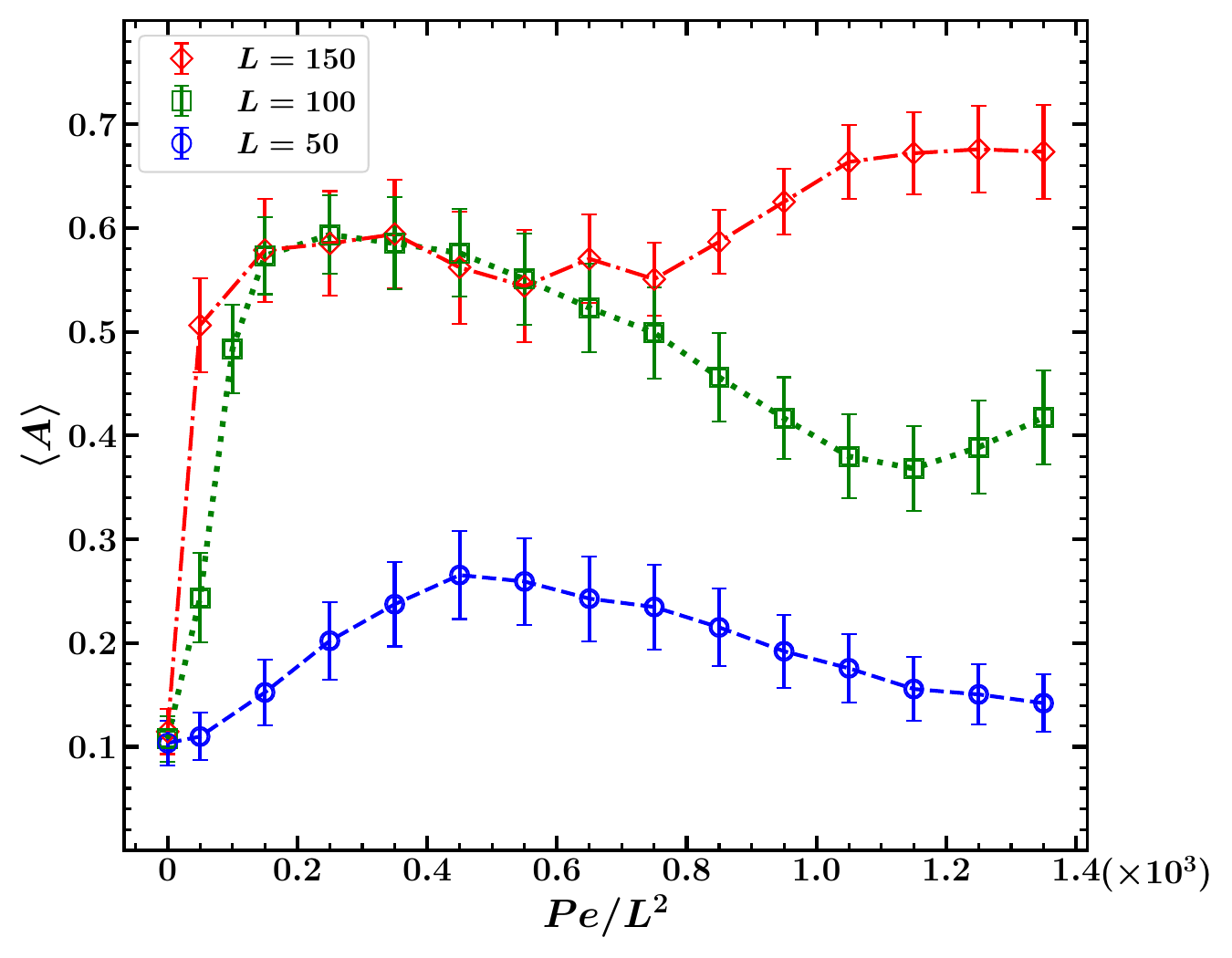}
\caption{\justifying The average asphericity $\langle A \rangle$ as a function of the P\'eclet number (Pe), with the bare processivity rate fixed at $\Omega = 0.5$ and persistence ratio $l_p/L = 0.5$. The polymer length is varied as $L = 50, 100, 150\,\sigma$}
\label{fig:asphericity_pe_varyL}
\end{figure}

 To assess how the conformational response depends on polymer length, Fig.~\ref{fig:asphericity_pe_varyL} shows the average asphericity \(\langle A\rangle\) as a function of activity for three ring lengths \(L=50\sigma\), \(100\sigma\), and \(150\sigma\) at fixed stiffness and processivity. Increasing \(L\) leads to a significant enhancement of \(\langle A\rangle\), indicating that longer rings can sustain larger shape distortions under motor-driven force. At \(L=50\sigma\), \(\langle A\rangle\) exhibits a distinctly non-monotonic variation with activity. As the length increases, \(\langle A\rangle\) increases more sharply at intermediate activity. \(\langle A\rangle\) remains elevated over a broader range of \(Pe\). This reflects the larger number of available deformation modes and the reduced effective cost of long-wavelength distortions. Notably, while the detailed shape of the \(\langle A\rangle(Pe)\) curve depends on \(L\), the qualitative conclusion that motor-driven forcing can generate substantial, activity-dependent deviations from circularity remains robust.

\section{MSD on Ring of Larger Length}

The MSD of a larger polymer ring exhibits behavior qualitatively similar to that observed for the \(L=50\sigma\) case. Since the relaxation timescale \(\tau\) depends on the polymer length, the corresponding value of \(\tau\) is different for the \(L=100\sigma\) system. 

\begin{figure}[h!tbp]
\centering
\includegraphics[width=\columnwidth]{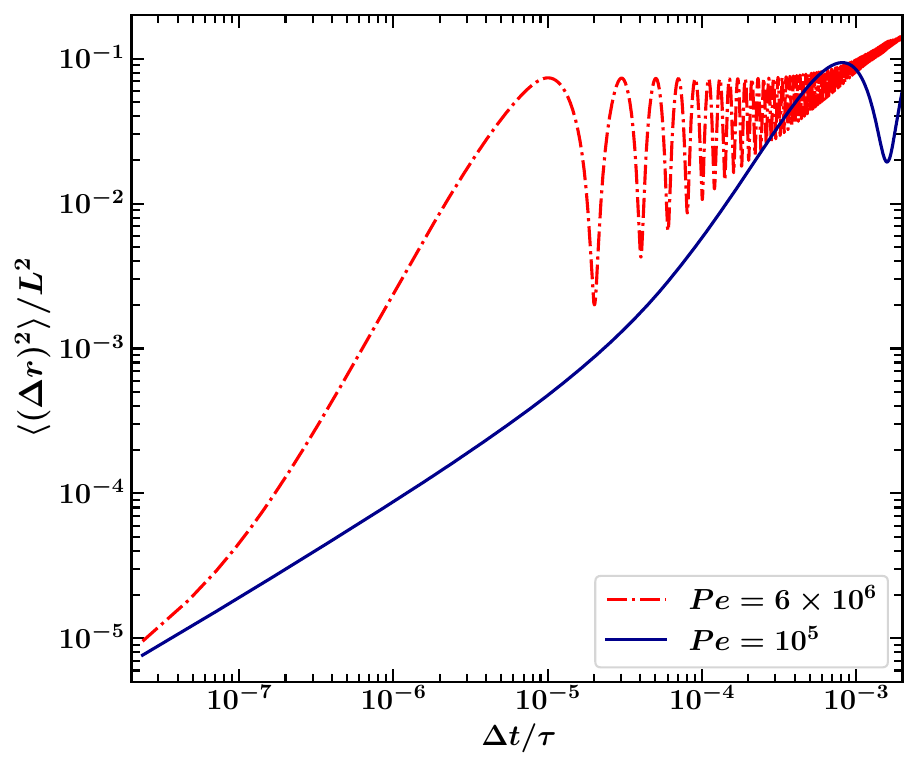}
\caption{\justifying Mean squared displacement (MSD) of a semiflexible polymer ring with length \(L = 100\sigma\) and persistence ratio \(l_p/L = 0.5\). \\}
\label{fig:MSD_L100}
\end{figure}

For small activity, \(Pe = 10^5\), the short-time behavior remains unchanged, with the MSD scaling as \(\Delta t^{0.7}\), similar to the \(L=50\sigma\) case. However, the oscillation amplitude appearing at \(\Delta t/\tau > 10^{-3}\) becomes significantly more pronounced for the longer ring. In this case, the crossover occurs from \(\Delta t^{0.7}\) to \(\Delta t^{1.5}\) (see Fig.~\ref{fig:MSD_L100}).

For larger activity, \(Pe = 6 \times 10^6\), strong oscillatory features appear in the MSD, reminiscent of coherent rotational motion of the polymer ring.

\section{Supplementary Movies}
\noindent
\textbf{Supplementary Movie I}:
\label{movie:1}
Time evolution of a semiflexible ring polymer on a motor-protein bed in the $x-y$ plane for $Pe =11.2\times 10^5, \Omega=0.1, l_p/L = 0.3$. Blue markers denote monomers; the red marker highlights a tagged monomer to aid visual tracking. The ring displays sustained rotational motion with relatively modest shape deformation over time.\\

\vspace{0.2cm}
\noindent
\textbf{Supplementary Movie II}:
\label{movie:2}
Same as Movie S1, but at higher motor processivity $\Omega=0.9$ ($Pe =11.2\times 10^5, l_p/L = 0.5$). Increased processivity leads to stronger, more heterogeneous shape distortions (elongation and corrugations) while the ring continues to undergo net rotation; the tagged monomer (red) illustrates the evolving orientation and deformation dynamics. \\

\section*{Author Contributions}
Conceptualization, A. K. Dasanna; methodology, S. Roy, A. Chaudhuri and A. K. Dasanna; investigation, S. Roy; writing -  S. Roy,  A. K. Dasanna and A. Chaudhuri; Software, S.Roy; supervision, A. Chaudhuri and A. K. Dasanna.

\section{Acknowledgements}
SR acknowledges TIFR Hyderabad, where part of the work was completed. Computing facilities at IISER Mohali and TIFR Hyderabad are acknowledged.

\section*{Declaration of Interests}
Authors declare no competing interests.

\section{Data availability}
The data associated with the figures is available at: \url{https://github.com/AnilBiophysics/RingPolymer_MotilityAssay_Sandip}

\bibliography{reference}

\end{document}